\documentclass[aps,twocolumn,showpacs,preprintnumbers,nofootinbib,prl,superscriptaddress,10pt]{revtex4-2}

\makeatletter
\def\l@subsubsection#1#2{}
\def\l@subsubsubsection#1#2{}
\makeatother

\usepackage{graphicx,amssymb,amsmath,amsthm,amsfonts,epsfig,epsf,fixmath}
\usepackage[usenames]{color}
\usepackage{epstopdf}

\usepackage{aas_macros}
\usepackage{bm}
\usepackage{comment}
\usepackage{dcolumn}
\usepackage{latexsym}
\usepackage{rotating}
\usepackage{longtable}

\usepackage{enumerate}
\usepackage{tensor,multirow}
\usepackage{url}
\usepackage[linktocpage]{hyperref}
\usepackage{soul}

\newcommand{\GSSI}{Gran Sasso Science Institute (GSSI), I-67100 L’Aquila, Italy}
\newcommand{\GranSasso}{INFN, Laboratori Nazionali del Gran Sasso, I-67100 Assergi, Italy}

\begin{document}
\title{Ranking Love numbers for the neutron star equation of state: The need for third-generation detectors}
\author{Costantino  Pacilio}
\affiliation{Dipartimento di Fisica, ``Sapienza'' Universit\`a di Roma, Piazzale
Aldo Moro 5, 00185, Roma, Italy}
\address{INFN, Sezione di Roma, Piazzale Aldo Moro 2, 00185, Roma, Italy}

\author{Andrea Maselli}
\address{\GSSI}
\address{\GranSasso}

\author{Margherita Fasano}
\affiliation{Dipartimento di Fisica, ``Sapienza'' Universit\`a di Roma, Piazzale
Aldo Moro 5, 00185, Roma, Italy}

\author{Paolo Pani}
\affiliation{Dipartimento di Fisica, ``Sapienza'' Universit\`a di Roma, Piazzale
Aldo Moro 5, 00185, Roma, Italy}
\address{Sezione INFN Roma1, Roma 00185, Italy}

\begin{abstract}

Gravitational-wave measurements of the tidal deformability 
in neutron-star binary coalescences can be used to infer the 
still unknown equation of state~(EoS) of dense matter above 
the nuclear saturation density. By employing a Bayesian-ranking 
test we quantify the ability of current and future gravitational-wave 
observations to discriminate among families of nuclear-physics 
based EoS which differ in particle content and ab-initio 
microscopic calculations.
While the constraining power of GW170817 is limited, 
we show that even twenty coalescences detected by 
LIGO-Virgo at design sensitivity are not enough to discriminate 
between EoS with similar softness but distinct microphysics.
However, just a single detection with a third-generation detector such 
as the Einstein Telescope or Cosmic Explorer will rule out several 
families of EoS with very strong statistical significance, and can 
discriminate among models which feature similar softness, hence
constraining the properties of nuclear matter to unprecedented levels.
\end{abstract}

\maketitle

%--------------------------------------------------------------------------------------
\noindent{{\bf{\em Introduction.}}}
%--------------------------------------------------------------------------------------
The equation of state~(EoS) of dense matter plays a crucial role 
in many astrophysical phenomena associated with neutron stars~(NSs) in different environments and dynamical regimes~\cite{Ozel:2010fw}. 
The electromagnetic~(EM) and gravitational-wave~(GW) signals emitted by isolated and (coalescing) binary NSs depend on the properties 
of the stellar structure and carry 
precious information on the nature of stellar cores where the density 
is much larger than the nuclear saturation point,
$\rho_0\approx 2.7\times 10^{14}$g/cm$^3$~\cite{Lattimer:2006xb,Ozel:2016oaf}. In this regime EoS models feature large uncertainties due to the complexity in describing 
strong interactions at densities where constituents other than 
nucleons may appear. This 
uncertainty reflects into a plethora of models with different 
particle content, featuring for example plain $npe\mu$ matter, hyperons, 
pion condensates, quarks, etc~\cite{Lattimer:2006xb}, and also predicting different macroscopic stellar properties, such as maximum mass, compactness, 
and tidal deformability~\cite{Ozel:2016oaf,PoissonWill,Chatziioannou:2020pqz}. This variety hampers our 
ability to uniquely characterise the behavior of nuclear matter in 
extreme conditions, and hence the NS structure.

Constraints on the EoS in the laboratory are limited by the density regime achievable by terrestrial 
experiments~\cite{symmetry:expt,russotto,Tsang,IAS,Brown,Zhang,danielewicz,Shlomo2006,Colo2008,Shlomo2006}. Major advances are expected to come from 
astrophysical observations, either from mass-radius measurements 
in the EM band~\cite{Demorest:2010bx,Antoniadis:2013pzd,Fonseca:2016tux,Cromartie:2019kug,
Ozel:2010fw,Steiner:2010fz,Guver:2013xa,Riley:2019yda,Miller:2019cac} 
or, more recently, from GW observations of 
binary NS mergers~\cite{TheLIGOScientific:2017qsa,Abbott:2018wiz,Abbott:2020uma,Baiotti:2016qnr}, where the EoS leaves an imprint in the latest stages of the inspiral and in the post-merger signal.
GW measurements of the tidal  
deformability of coalescing NS binaries~\cite{Flanagan:2007ix,Hinderer:2007mb} provide a new 
tool to probe the behavior of matter at densities above
$\rho_0$~\cite{TheLIGOScientific:2017qsa,Hinderer:2010ih,Baiotti:2010xh,Baiotti:2011am,Vines:2011ud,Pannarale:2011pk,Vines:2010ca,Lackey:2011vz,Maselli:2013rza,Lackey:2013axa,
  Favata:2013rwa,DelPozzo:2013ala,Lackey:2014fwa} (see~\cite{GuerraChaves:2019foa,Chatziioannou:2020pqz} for recent reviews). 
The landmark detection of GW170817 has already ruled out 
very stiff EoS which predict large tidal deformabilities~\cite{TheLIGOScientific:2017qsa, Abbott:2018wiz}.  
Moreover, the detection of an EM counterpart to GW170817 has motivated  several
multimessenger analyses aimed at providing joint GW-EM 
constraints~\cite{Annala:2017llu,Margalit:2017dij,Radice:2017lry,Bauswein:2017vtn,Lim:2018bkq,Most:2018hfd,Carson:2018xri,De:2018uhw,Annala:2019puf,Raaijmakers:2019dks,Capano:2019eae,Miller:2019nzo,Kumar:2019xgp,Fasano:2019zwm,Guven:2020dok,Traversi:2020aaa,Landry:2020vaw,Dietrich:2020efo,Zimmerman:2020eho,Silva:2020acr,Al-Mamun:2020vzu,Sabatucci2020,Maselli:2020uol} 
(see~\cite{Baiotti2019,Chatziioannou:2020pqz,Ozel:2016oaf} 
for some reviews). 

The majority of these approaches interpreted 
constraints on the tidal deformability using phenomenological EoS, 
which map wide samples of models in terms of a relatively 
small set of parameters~\cite{Read:2008iy,Lindblom:2010bb,Hebeler:2013nza,Greif:2018njt,Hebeler:2015hla,Tews:2018kmu,Biswas:2020puz}, or synthetic EoS~\cite{Landry:2018prl}.
While flexible, these models lack the 
description of the microphysical content which otherwise 
characterises ab-initio, nuclear-physics based EoS. 
In this work we pursue a complementary approach and try to answer 
the following question: given a set of nuclear-physics based cold EoS --~which differ in
the particle content and in the ab-initio microscopic calculations~--
what is the one that is mostly favored (in a rigorous statistical sense) by current and future observations?

%%%%%%%%%%%%%%%%%%%%%%%%%%%%%%%%%%%%%%%%%%%%%%%%%%%%%%%
\begin{figure}[th]
\centering
\includegraphics[width=0.5\textwidth]{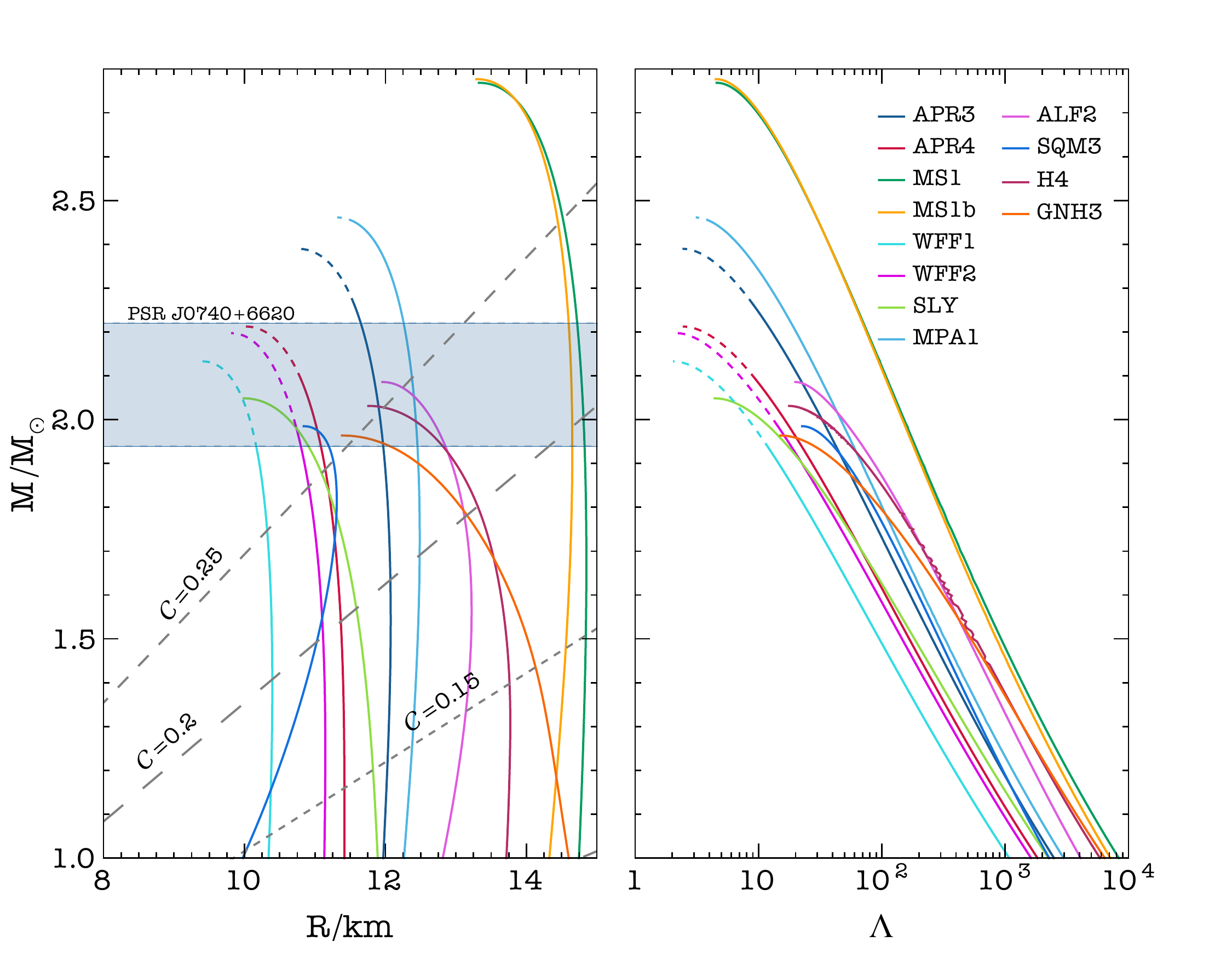}
\caption{Mass-radius and mass-tidal deformability diagrams 
for the EoS considered in the Bayesian analysis. The blue 
band on the left panel corresponds to the most massive pulsar 
observed in the EM band ($M=2.08^{+0.07}_{-0.07}\,M_\odot$~\cite{Fonseca:2021wxt}), 
while dashed lines identify configurations with fixed 
compactness ${\cal C}=M/R$. Solid (dashed) curves 
correspond to stellar configurations with the 
speed of sound at the center smaller (larger) 
than the speed of light.}
\label{fig:MR} 
\end{figure}
%%%%%%%%%%%%%%%%%%%%%%%%%%%%%%%%%%%%%%%%%%%%%%%%%%%%%%%

In order to address this problem,
we perform a hierarchical Bayesian test that --~given a set of GW data on the binary masses and tidal deformability~-- ranks different models of dense 
matter according to their statistical evidence.
We first apply this method to the real data of GW170817, confirming that the constraining power of this event is limited to excluding only very stiff EoS~\cite{LIGOScientific:2019eut}. We then extend this approach to a near-future scenario, using current interferometers at design  
sensitivity and stacking multiple binary NS observations 
characterised by different masses and distances~\cite{DelPozzo:2013ala,Lackey:2014fwa}. 
Our results show that the sensitivity of the advanced 
LIGO/Virgo interferometers is not sufficient to 
resolve the degeneracy between EoS featuring similar 
softness.
We therefore apply, for the first time, this Bayesian analysis 
to the Einstein Telescope~(ET), a proposed third-generation 
ground-based GW observatory~\cite{Punturo:2010zz,PhysRevD.91.082001,Evans:2016mbw,Essick:2017wyl,Hild:2010id,Sathyaprakash:2019yqt,Maggiore:2019uih}. 
In this case, we found that even a single ordinary detection would rule out several classes of EoS and is sufficient to discriminate among nuclear-matter models with similar softness. Furthermore, just stacking few 
detections would be sufficient to pinpoint a single EoS with decisive statistical evidence.

%---------------------------------------------------------------
\noindent{{\bf{\em EoS catalog and dataset simulations.}}}
%---------------------------------------------------------------
We consider 12 state-of-the-art EoS which can be classified into three broad families depending on their matter content: (i)~plain $npe\mu$ nuclear matter --- APR3, APR4, SLY, MPA1, MS1, MS1b, WFF1, WFF2~\cite{Akmal:1998cf,Douchin:2001sv,Muther:1987xaa,Mueller:1996pm,Wiringa:1988tp}; (ii)~ models with hyperons --- GNH3, H4~\cite{Glendenning:1984jr,Lackey:2005tk}; and (iii) hybrid EoS with mixtures of nucleonic and quark matter --- ALF2, SQM3~\cite{Alford:2004pf,Prakash:1995uw}. Naming conventions follow~\cite{Lattimer:2000nx,Ozel:2016oaf}. This ensemble of EoS encompasses a wide range of stiffness. For a reference mass $M=1.4 M_\odot$, they predict compactness in the range ${\cal C}=M/R\in(0.14,0.20)$ and dimensionless tidal deformabilities in the range $\Lambda\in(151,1377)$, see Fig.~\ref{fig:MR} and Table~\ref{table_eos}.

The EoS have been selected to be compatible with J$0740$+$6620$~\cite{Fonseca:2021wxt}, the most massive pulsar observed to date ($M=2.08^{+0.07}_{-0.07}\,M_\odot$ at $68.3\%$ confidence level). In particular, all the considered EoS have a maximum mass  above the ($2\sigma$) lower bound $1.94 M_\odot$ and subluminal sound speed in the relevant mass range. For some EoS, this restricts the range of allowed configurations (e.g., WFF1 marginally satisfies the causality condition).

%TTTTTTTTTTTTTTTTTTTTTTTTTTTTTTTTTTTTTTTTTTTTTTTTTTTTTTTTTT
\begin{table}[th]
\centering
\begin{tabular}{cccc}
\hline
\hline
EoS & ${\rm family}$ & ${\rm particles}$ & $\Lambda_{1.4}$\\
\hline
ALF2 & nmbt+bag &$npe\mu+Q$ & 754\\
APR3 & nmbt & $npe\mu$ & 390\\
APR4 & nmbt & $npe\mu$ & 261\\
GNH3 & mft & $npe\mu+H$ & 866\\
H4 & mft & $npe\mu+H$ & 897\\
MPA1 & mft & $npe\mu$ & 487\\
MS1 & mft & $npe\mu$ & 1377\\
MS1b & mft & $npe\mu$ & 1250\\
SLY & mft & $npe\mu$ & 297\\
SQM3 & mft+bag & $npe\mu+H+Q$ & 432\\
WFF1 & nmbt & $npe\mu$ & 151\\
WFF2 & nmbt & $npe\mu$ & 229\\
\hline
\hline
\end{tabular}
\caption{List of the selected EoS with the corresponding calculation 
methods (family), particle content, and dimensionless tidal deformability 
at the reference mass $M=1.4 M_\odot$. The families are distinguished in: nuclear many body~(nmbt) calculations and mean-field theory (mft)
(see~\cite{benhar2020nuclear} for a review on EoS calculations). In 
the ALF2 and SQM3 EoS the quark~(Q) content is modelled according to the 
MIT bag model, while the GNH3, H4 and SQM3 EoS include hyperons~(H).}
\label{table_eos}
\end{table}
%TTTTTTTTTTTTTTTTTTTTTTTTTTTTTTTTTTTTTTTTTTTTTTTTTTTTTTTTTT

Besides analyzing the single GW170817 binary NS event, we simulate 
two selected catalogs of binary NS events consisting of 20 GW sources 
(see Appendix). The selected masses are drawn uniformly within 
$(1.2,1.6)\,M_\odot$, which is compatible with the mass range inferred 
for GW170817, and luminosity distance $d_L$ drawn uniformly in comoving 
volume with $60\leq d_L/{\rm Mpc}\leq 210$. 
We emphasize that, given the large number of binary-NS events expected in the third-generation era~\cite{Mapelli:2018wys} one can restrict to a subset of optimal observations, e.g. including only the loudest events with relatively small component masses, which provide the best constraints on the EoS.
The injected signals in the two 
catalogs assume the EoS APR4 and ALF2, respectively, as prototypes of 
soft and stiff nuclear matter.

We use the \texttt{IMRPhenomPv2\_NRTidal} 
model~\cite{Dietrich:2017aum,Dietrich:2018uni} GW waveform 
template. We inject nonspinning binaries, and we recover 
the component spins imposing a low-spin prior 
$\chi_{1,2}\in[-0.05,0.05]$ and assuming spins are 
(anti-)aligned.
To help comparison between the events, we fix the same sky location 
and inclination for all sources, avoiding particularly optimistic or 
pessimistic choices. 
We inject 64-second long waveforms into a zero-noise configuration as described 
in~\cite{Wade:2014vqa}, either for a network composed by the LIGO Hanford, 
LIGO Livingston, and Virgo detectors at design sensitivity~\cite{Pitkin:2011yk}, 
or for the future third-generation interferometer Einstein Telescope 
in its ET-D configuration~\cite{Hild:2010id}. 
We checked that our results remain valid also when using a random realization of the detector noise.

For a given simulated observation we reconstruct the posterior 
probability distribution of the waveform parameters using the 
publicly available BILBY code, a Bayesian inference library for 
GW astronomy~\cite{Ashton:2018jfp,Romero-Shaw:2020owr}. 
We use analytic marginalization procedures for the binary 
orbital phase, luminosity distance, and time of coalescence, as 
described in~\cite{Romero-Shaw:2020owr}.
We marginalize on the inferred 
posterior probability distribution to extract the joint probability 
function $\mathcal{P}({\cal M},\eta,\tilde{\Lambda})$ for the binary 
chirp mass ${\cal M}$, symmetric mass ratio $\eta$, and effective tidal 
deformability~\cite{Flanagan:2007ix}
\begin{equation}
    \label{eq:lambda:tilde}
    \begin{split}
    \tilde{\Lambda}=&\frac{16}{13}\left[\frac{\left(m_1+12\,m_2\right)m_1^4\Lambda_1}{(m_1+m_2)^5}+1\leftrightarrow2\right]\,.
    \end{split}
\end{equation}
For a given EoS, $\tilde\Lambda$ depends only on the two 
source-frame masses $m_1$ and $m_2$ or, 
equivalently, on ${\cal M}$ and $\eta$.

%---------------------------------------------------------------
\noindent{{\bf{\em Bayesian methods.}}}
%---------------------------------------------------------------
%
Given the data $\mathcal{D}$ from a GW event compatible with a coalescing NS binary, the degree of belief that the two NSs obey a given EoS can be quantified by the evidence~\cite{Raaijmakers:2019dks}
\begin{equation}
\begin{split}
    \label{eq:evidence}
    \mathcal{Z}(\mathcal{D}|{\rm EoS})=&\int_a^b dp^{(1)}\int_a^b dp^{(2)}\,\mathcal{P}({\cal M},\eta,\tilde{\Lambda}|\mathcal{D})\\
    &\times\mathcal{P}(p^{(1)}|{\rm EoS})\,\mathcal{P}(p^{(2)}|{\rm EoS})\,,
\end{split}
\end{equation}
where $p^{(1)}$ and $p^{(2)}$ are the central pressures of the two NSs.
For any given EoS, there is a deterministic mapping between the central pressures and the waveform parameters,  $\{p^{(1)},p^{(2)}\}\to\{{\cal M},\eta,\tilde{\Lambda}\}$, and in the above equation $\{{\cal M},\eta,\tilde{\Lambda}\}$ are evaluated as functions of $\{p^{(1)},p^{(2)}\}$.

The priors on the central pressures are uniform distributions within $p^{(i)}\in [a,b]$, where $a=p_{\rm min}\simeq1.21\times10^{34}{\rm dyne/cm}^2$ and $b=p_{\rm max}$ corresponds, for a given EoS, to the value of the pressure which yields the maximum mass configuration compatible with causality.

The calculation of the evidence in Eq.\,\eqref{eq:evidence} can be largely simplified 
using the fact that the chirp mass of NS binaries is measured with exquisite 
precision~\cite{Wade:2014vqa}, since these sources perform several cycles in band. 
(For example, the chirp mass of GW$170817$ was measured with $\approx 0.1\%$ precision, 
much better than any other intrinsic parameter~\cite{TheLIGOScientific:2017qsa}.) 
Therefore, in Eq.\,\eqref{eq:evidence}, we can fix ${\cal M}$ to its median inferred value 
${\cal M}_\star$.
Note that an accurate measurement of the source-frame masses solely from GWs can be 
hindered by the well-known degeneracy between the inclination angle and the luminosity 
distance~\cite{Usman:2018imj,Chen:2018omi}, which may induce potential biases in the redshift measurement. To resolve this degeneracy we assume that the redshift of the selected events is known (e.g. if independently measured by an EM counterpart as in GW170817~\cite{Abbott:2018wiz}). Thus, we fix $\mathcal{M}_\star=\mathcal{M}_\star^{\rm det}/(1+z)$, where $\mathcal{M}_\star^{\rm det}$ is the median of the inferred distribution of the detector-frame chirp mass, and $z$ is the injected value of the redshift.
We also verified that our analysis is not significantly affected by shifting $z$ away 
from its injected value by $\pm 10\%$, which is very conservative since it corresponds to the accuracy in $z$ as measured from the GW170817 EM counterpart~\cite{Abbott:2018wiz}.

Following~\cite{Raaijmakers:2019dks}, the conditional probability $\mathcal{P}(\eta,\tilde{\Lambda}|\mathcal{M}_\star,\mathcal{D})$ can be replaced by the marginalized probability $\mathcal{P}(\eta,\tilde{\Lambda}|\mathcal{D})$ to a very good approximation, and the evidence reduces to
\begin{eqnarray}
    \label{eq:evidence:2}
    \mathcal{Z}(\mathcal{D}|{\rm EoS})=&&\int_a^b dp^{(1)}\,\mathcal{P}(\eta(p^{(1)},p^{(2)}_\star),\tilde{\Lambda}(p^{(1)},p^{(2)}_\star)|\mathcal{D})\nonumber\\
    &\times&\mathcal{P}(p^{(1)}|{\rm EoS})\,\mathcal{P}(p^{(2)}_\star|{\rm EoS})\ ,
\end{eqnarray}
where $p_\star^{(2)}$ is the solution (if it exists) of $\mathcal{M}(p^{(1)},p^{(2)}_\star)=\mathcal{M}_\star$.
The above equations assume that the EoS configurations are sampled uniformly with respect to the central pressures. However, one could have equally used any monotonic function of the pressure. In particular, we opt for sampling the EoS uniformly with respect to $\log_{10}(p^{(1)})$ and change the integral in Eq.~\eqref{eq:evidence:2} accordingly.

%%%%%%%%%%%%%%%%
\begin{figure}[t]
\centering
\includegraphics[width=0.4\textwidth]{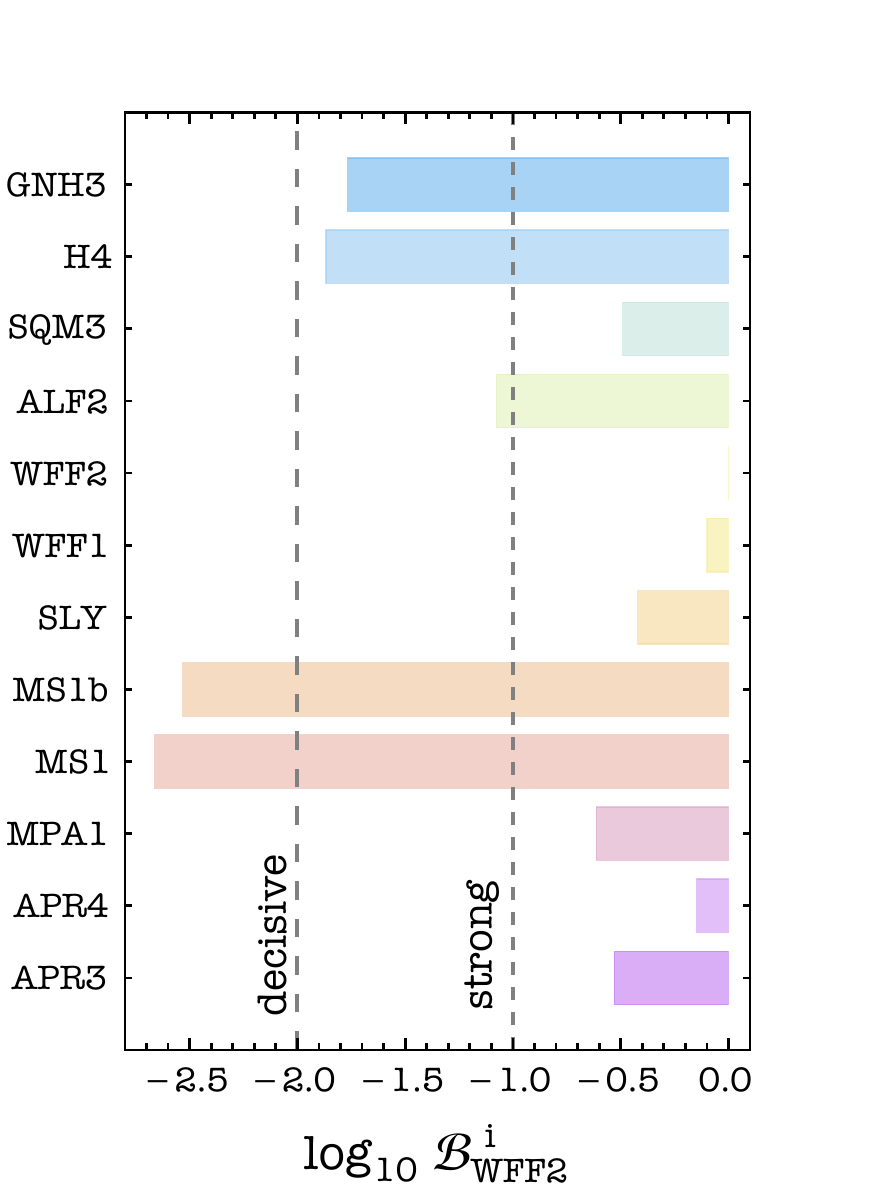}
\caption{Bayes factors for different EoS models computed for GW170817 and normalized with respect to the 
EoS with maximum evidence in the catalogue~(WFF2). Vertical 
dashed lines identify the threshold above which the Bayes 
factor provide a strong and decisive evidence in favor of 
WFF2.}
\label{fig:LVC} 
\end{figure}
%%%%%%%%%%%%%%%%%%%%%%%%%%%%%%%%%%%%%%%%%%%%%%%%%%%%%%%

We can use the Bayes factor,
\begin{equation}
\label{eq:bfactor}
\mathcal{B}^1_2=\frac{\mathcal{Z}(\mathcal{D}|{\rm EoS}_1)}{\mathcal{Z}(\mathcal{D}|{\rm EoS}_2)}\,,
\end{equation}
to express the relative odds of two EoS given the data $\mathcal{D}$, assuming equal priors on the EoS, $\mathcal{P}({\rm EoS}_1)=\mathcal{P}({\rm EoS}_2)$.

The previous discussion can be easily extended to the case of stacked observations $\vec{\mathcal{D}}=\{\mathcal{D}_1\dots\mathcal{D}_n\}$.
After $n$ observations the relative odds will be updated by the cumulative Bayes factor,
\begin{equation}
    \label{eq:odds:2}
    \mathcal{B}^1_2=\prod_{k=1}^n\frac{\mathcal{Z}(\mathcal{D}_k|{\rm EoS}_1)}{\mathcal{Z}(\mathcal{D}_k|{\rm EoS}_2)}\,.
\end{equation}
The main quantity of interest is the cumulative logarithmic Bayes factor, $\log_{10}{\cal B}^i_T$,
between a candidate EoS$_i$ and a benchmark EoS$_T$ after $n$ GW detections. We adopt the Kass-Raftery criterion~\cite{kass1995bayes} and decisively exclude ${\rm EoS}_i$ with respect to ${\rm EoS}_T$ when $\log_{10}\mathcal{B}_T^i<-2$.

%--------------------------------------------------------------------
\noindent{{\bf{\em Results.}}}
%--------------------------------------------------------------------
%
We start by applying this method to real data, using GW170817~\cite{Abbott:2018wiz,TheLIGOScientific:2017qsa}, the only 
binary NS GW event --~among those detected so far by LIGO and Virgo~\cite{LIGOScientific:2018mvr,Abbott:2020niy}~-- that provided an accurate measurement of the tidal deformability~\cite{Abbott:2018wiz,Abbott:2018exr}.
Figure~\ref{fig:LVC} shows the Bayes factors of different EoS in the 
catalog normalized with respect to the EoS with maximum evidence, 
which turns out to be WFF2. The evidence against other EoS is weak 
in most cases, except for GNH3 and H4, and especially for MS1 and 
MS1b which are decisively excluded according to the Kass-Raftery
scale. This is in agreement with the fact that MS1 and MS1b are the 
stiffest EoS in our catalog and therefore the easiest to rule out with GW170817~\cite{Abbott:2018wiz,Abbott:2018exr,Bauswein:2017vtn,Annala:2017llu,
Most:2018hfd,Harry:2018hke,De:2018uhw,LIGOScientific:2019eut}. Likewise, EoS stiffer than MS1 and MS1b are even more disfavored by GW170817.

\begin{figure}[th]
\centering
\includegraphics[width=0.5\textwidth]{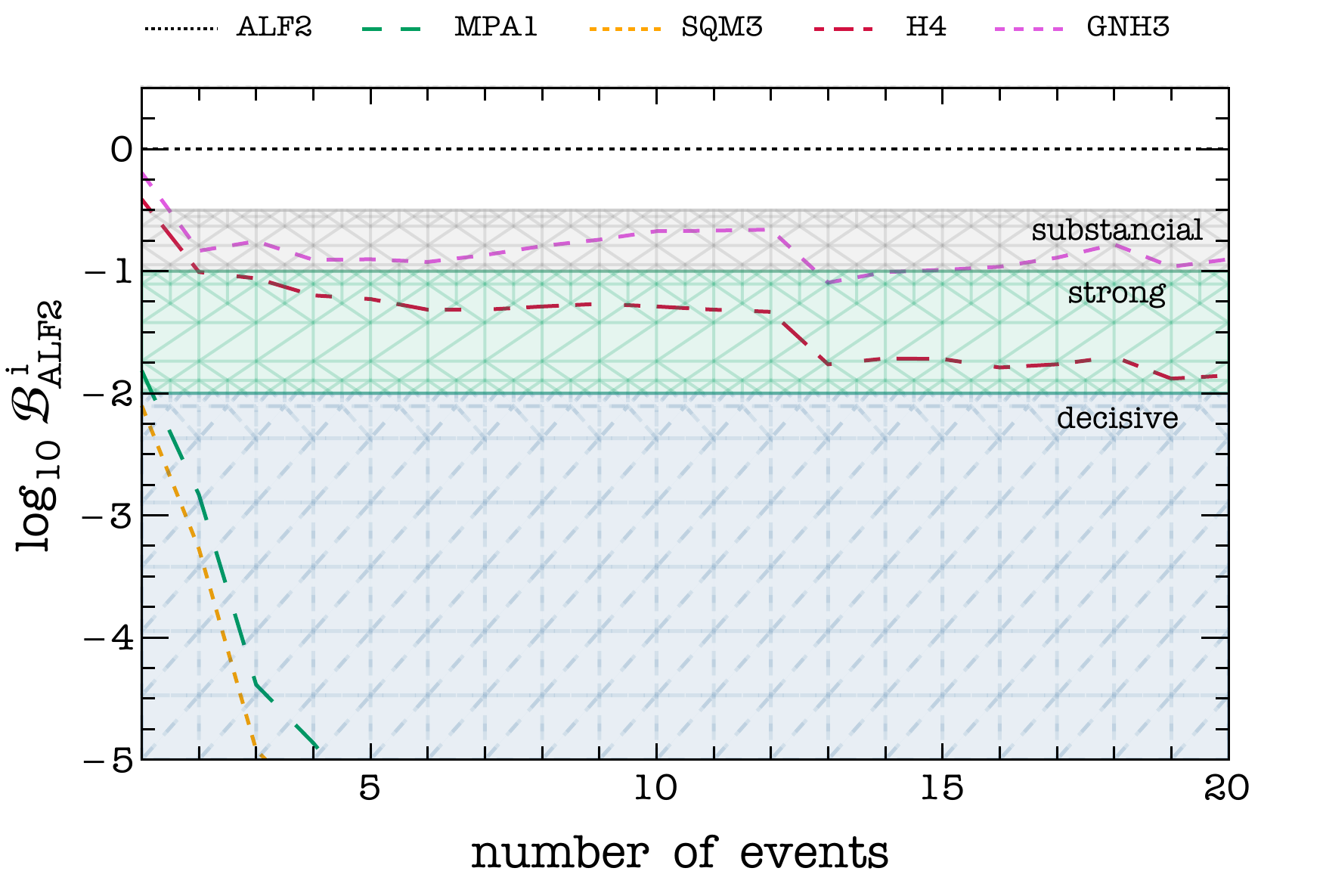}
\includegraphics[width=0.5\textwidth]{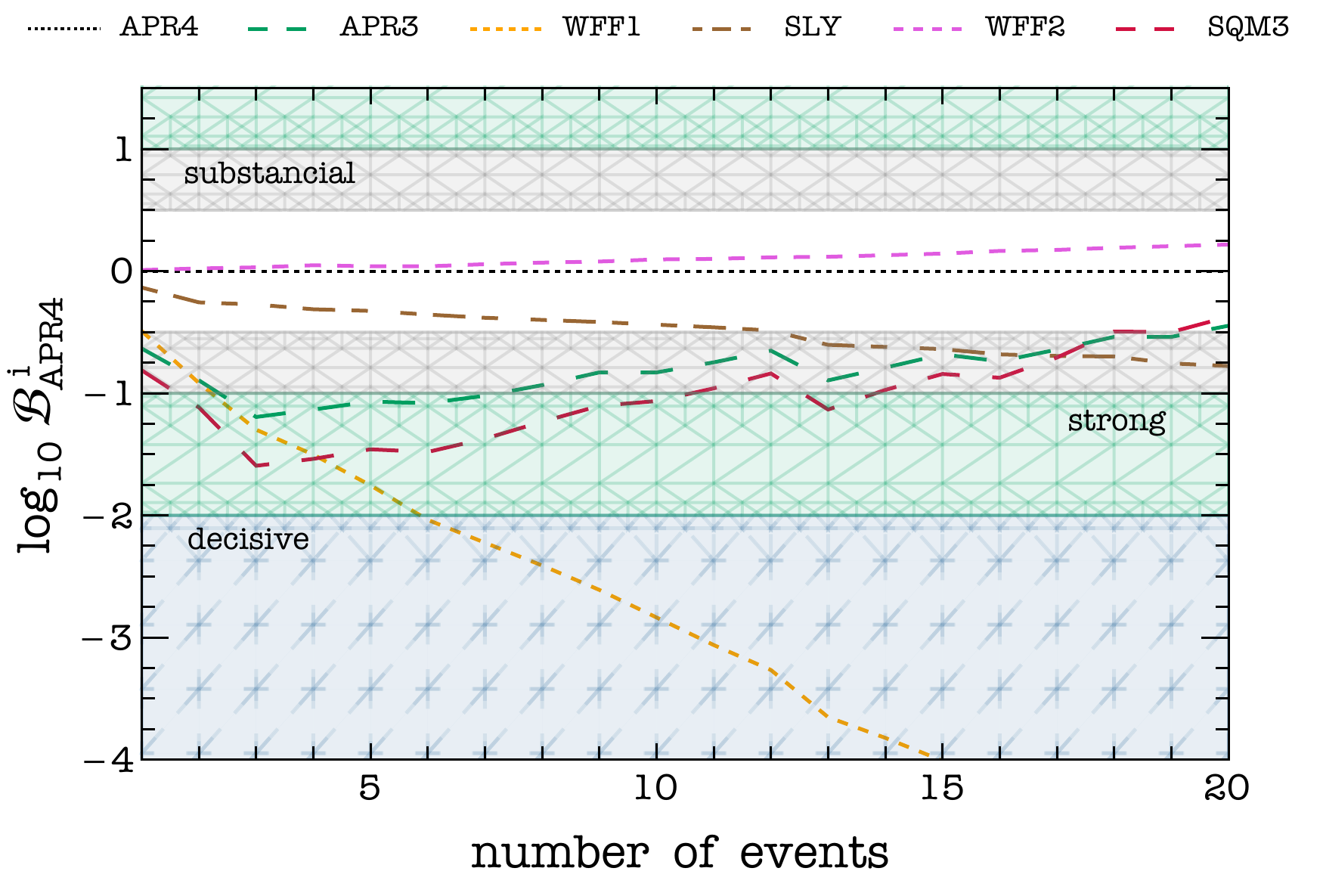}
\caption{Evolution of the EoS Bayes factor with the number of 
events for the LIGO-Virgo network at design sensitivity. 
Top and bottom panels refer to the ALF2 (stiff EoS) and AP4 (soft EoS) injections, 
respectively. In each panel, the quantity $\log_{10}{\cal B}^i_T$ 
is normalized with respect to the injected EoS. Shaded bands mark 
the boundaries of the evidence criteria according to the Kass-Raftery
scale~\cite{kass1995bayes}. 
In particular, $\log_{10}{\cal B}^i_T<-2$ indicates decisive 
unfavorable evidence.}
\label{fig:bf_HLV}
\end{figure}

Stronger constraints and statistical evidence can be obtained from 
accumulating more detections~\cite{DelPozzo:2013ala,Lackey:2014fwa}. 
In Fig.~\ref{fig:bf_HLV} we show the Bayes factor as a function of the 
number of randomly chosen events detected by the advanced LIGO-Virgo 
network at design sensitivity and assuming the real EoS is either: i) 
relatively stiff (ALF2, top panel) or ii) relatively soft (APR4, bottom 
panel). In each panel we show only the subset of EoS with the highest 
Bayes factors, whereas the other EoS are easier to rule out. 
In both cases it is challenging 
to rule out EoS with stiffness similar to the reference 
one even after 20 detections (this is more evident for a 
soft model such as APR4, shown in the bottom panel).
This analysis shows, in a clear and statistically robust way, that while several LIGO-Virgo detections at design sensitivity could discriminate among some stiff EoS (e.g. ALF2 versus MPA1 and SQM3) and between some soft and stiff models~\cite{DelPozzo:2013ala}, they remain inconclusive, since the sensitivity is not enough to discriminate among wide classes of EoS with similar stiffness.

%%%%%%%%%%%%%%%%%%%%%%%%%%%%%}%%%%%%%%%%%%%%%%%%%%%%%%%%
%%%%%%%%%%%%%%%%%%%%%%%%%%%%%%%%%%%%%%%%%%%%%%%%%%%%%%%

\begin{figure}[!htbp]
\centering
\includegraphics[width=0.5\textwidth]{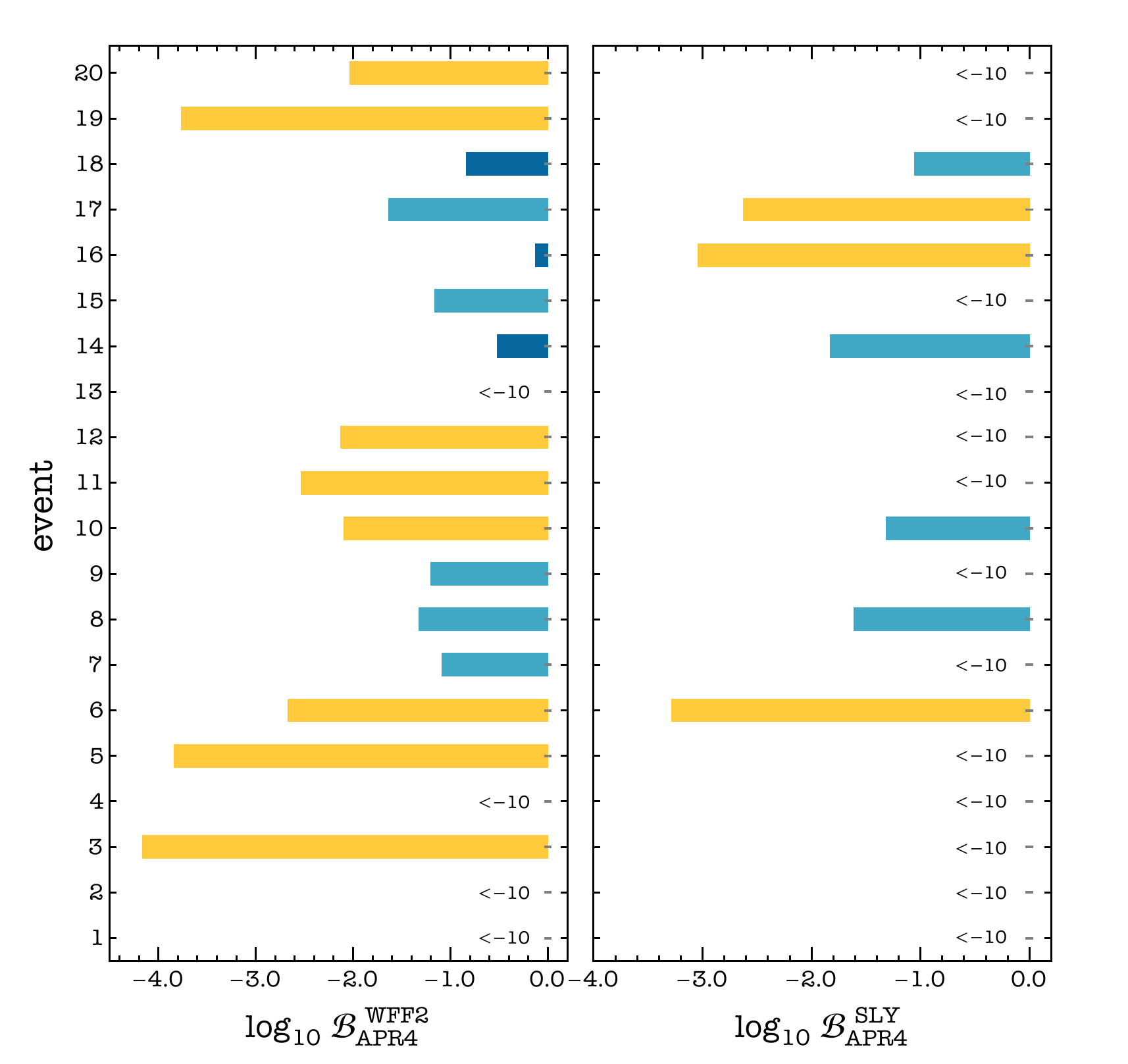}
\caption{Bayes factors for simulated observations with ET, 
relative to the injected EoS APR4, for WFF2 and SLY. 
The remaining set of ten EoS yield $\log_{10} {\cal B}_{\rm APR4}^i<-10$ for all events, and they are not shown in the plot.}
\label{fig:bf_ET}
\end{figure}

The latter conclusion motivates to forecast a similar analysis in the 
era of third-generation 
GW detectors~\cite{Punturo:2010zz,PhysRevD.91.082001,Evans:2016mbw,Essick:2017wyl,Hild:2010id,Sathyaprakash:2019yqt,Maggiore:2019uih}. The situation here 
drastically changes, as shown in Fig.~\ref{fig:bf_ET}. We simulated the 
same 20 detections with ET, by assuming the conservative case of an underlying APR4 
EoS, as in the bottom panel of Fig.~\ref{fig:bf_HLV}. For each event, we 
plot the Bayes factors normalized by the injected EoS and we only show 
those EoS which have nonvanishing evidence 
($\log_{10}{\cal B}_{\rm APR4}^i>-10$) for at least one event. The fact that most EoS have negligible evidence is a consequence of the much higher sensitivity 
of the ET detector, and it allows us to exclude \emph{all but a couple} of 
EoS of our dataset (namely WFF2 and SLy, which feature a tidal deformability similar to APR4) with only a single observation.

Even in the most pessimistic case, in which a single observation 
is not enough to exclude a given EoS, stacking two/three 
detections would allow us to decisively exclude {\rm all} EoS in 
the catalog other than the reference one.
Even stronger conclusions apply to the case in which the reference EoS are stiff (as for ALF2): in this case all the other EoS in our catalog are decisively excluded for any single event. 

Thus, at variance with advanced LIGO/Virgo, ET will be able to distinguish 
among EoS with similar softness, and also among EoS families featuring 
different microphysical properties (see Table~\ref{table_eos}). For example, a 
single ET detection of any of the $20$ events considered in our catalog 
would be sufficient to exclude APR3 relative to APR4 ($\log_{10}{\cal B}^{\rm APR3}_{\rm APR4}<-10$). These two EoS feature the same particle content but differ in the description of the nucleon interaction.

%--------------------------------------------------------------------
\noindent{{\bf{\em Conclusions.}}}
%--------------------------------------------------------------------
We proposed a robust Bayesian-ranking 
test to discriminate among families of ab-initio nuclear EoS using GW observations. We applied this test to GW170817, which very mildly favors a relatively soft, standard $npe\mu$ EoS (WFF2), although its power in ruling out EoS with similar stiffness is limited. 
Furthermore, we showed that near-future observations will not be conclusive: even $20$ NS binary detections with LIGO-Virgo at design sensitivity will not be able to distinguish among well-motivated nuclear models.

On the other hand, a single detection by ET will rule out with decisive statistical evidence most of the EoS, including those with comparable softness. In addition, just a few combined detections can be sufficient to robustly identify the best-fit EoS within a catalogue, hence constraining the particle content of nuclear matter at ultrahigh density. 
The same conclusion would apply assuming 
that binaries are observed by the proposed Cosmic 
Explorer~\cite{Evans:2016mbw,Essick:2017wyl}, which features a 
noise curve similar to that of ET-D at high frequencies, 
where tidal effects contribute more to the GW signal. 
Joint detections by ET and Cosmic Explorer would further 
strengthen our results.

Measuring the masses and tidal deformabilities from multiple events would allow us to quantify the faithfulness of the best-fit EoS, e.g. by looking for inconsistencies between the best-fit predictions and the data in the $\Lambda-M$ plane (see Fig.~\ref{fig:MR}), in case the ``true'' EoS is not in the dataset.

A further advantage of our approach based on a ranking test among nuclear-physics based EoS is that it can be straightforwardly extended to accommodate other measurements by combining the likelihoods of different models. It would be interesting to extend our analysis in this direction by combining future GW observations with EM ones~\cite{Raaijmakers:2019dks,Zimmerman:2020eho}, or with post-merger signals~\cite{Baiotti:2016qnr}.

%%%%%%%%%%%%%%%%%%%%%%%%%%%%%%%%%%%%%%%%%%%%%%%%%%%%%%%
%%%%%%%%%%%%%%%%%%%%%%%%%%%%%%%%%%%%%%%%%%%%%%%%%%%%%%%

%---------------------------------------------------------------
\noindent{{\bf{\em Acknowledgments.}}}
%---------------------------------------------------------------
%\begin{}
We are indebted to Swetha Bhagwat, Omar Benhar, 
Marica Branchesi, and Valeria Ferrari for discussions and comments on the manuscript. 
Numerical calculations have been made possible through 
a CINECA-INFN agreement, providing access to resources on 
MARCONI at CINECA. We acknowledge financial support provided 
under the European Union's H2020 ERC, Starting Grant agreement 
no.~DarkGRA--757480. We also acknowledge support under the MIUR 
PRIN and FARE programmes (GW-NEXT, CUP:~B84I20000100001), and 
from the Amaldi Research Center funded by the MIUR 
program ``Dipartimento di Eccellenza'' (CUP: B81I18001170001).

%\begin{comment}
\appendix
\noindent{{\bf{\em Appendix.}}}
In Table~\ref{table_source} we list the masses and distances of the 20 binary NS sources considered in the main text. The masses were drawn uniformly in the range $(1.2,1.6)\,M_\odot$, whereas the luminosity distance $d_L$ was drawn uniformly in comoving 
volume with $60\leq d_L/{\rm Mpc}\leq 210$.
For each event, we also show the 68\% confidence intervals around the median for the tidal deformability inferred by projected LIGO-Virgo network and ET observations.

\begin{table*}%[ht]
\centering
\begin{tabular*}{\textwidth}{cccccc|ccccc}
%\begin{tabular*}{cccccc|ccccc}
\hline
\hline
${\rm event}$ & $m_1/M_\odot$ & $m_2/M_\odot$ & $d_\textnormal{L}/{\rm Mpc}$ 
& $\tilde{\Lambda}_\textnormal{APR4}$ 
& $\tilde{\Lambda}_\textnormal{ALF2}$ 
& $\tilde{\Lambda}^\textnormal{HLV}_\textnormal{APR4}$ 
& $\tilde{\Lambda}^\textnormal{HLV}_\textnormal{ALF2}$ 
& $\tilde{\Lambda}^\textnormal{ET}_\textnormal{APR4}$ \\ %$\tilde{\Lambda}^\textnormal{HLV}_\textnormal{APR4}$ & %$\tilde{\Lambda}^\textnormal{HLV}_\textnormal{ALF2}$\\
\hline
1 & 1.51 & 1.37 & 61 & 220 & 643 
& $221^{+71}_{-62}$ & $627^{+79}_{-81}$ & $220^{+4}_{-4}$\\
2 & 1.30 & 1.27 & 107 & 442 & 1200 
& $476^{+257}_{-159}$ & $1132^{+217}_{-218}$ & $441^{+8}_{-9}$\\
3 & 1.57 & 1.55 & 64 & 128 & 395 
& $138^{+53}_{-46}$ & $386^{+62}_{-60}$ & $127^{+3}_{-3}$\\
4 & 1.23 & 1.23 & 179 & 573 & 1510 
& $1085^{+880}_{-544}$ & $1519^{+766}_{-532}$ & $563^{+12}_{-12}$\\
5 & 1.48 & 1.25 & 165 & 313 & 873 
& $683^{+646}_{-351}$ & $968^{+527}_{-329}$ & $312^{+12}_{-9}$\\
6 & 1.54 & 1.28 & 132 & 257 & 730 
& $386^{+350}_{-169}$ & $745^{+284}_{-204}$ & $257^{+8}_{-8}$\\
7 & 1.47 & 1.40 & 167 & 223 & 655 
& $549^{+561}_{-302}$ & $741^{+474}_{-264}$ & $216^{+9}_{-9}$\\
8 & 1.48 & 1.36 & 182 & 240 & 697 
& $693^{+671}_{-391}$ & $869^{+613}_{-345}$ & $239^{+12}_{-13}$\\
9 & 1.33 & 1.33 & 204 & 359 & 1000 
& $1049^{+889}_{-573}$ & $1247^{+826}_{-524}$ & $349^{+13}_{-14}$\\
10 & 1.58 & 1.42 & 133 & 169 & 506 
& $275^{+274}_{-127}$ & $518^{+220}_{-147}$ & $168^{+7}_{-7}$\\
11 & 1.39 & 1.29 & 182 & 345 & 962 
& $809^{+729}_{-427}$ & $1078^{+665}_{-394}$ & $342^{+11}_{-12}$\\
12 & 1.35 & 1.31 & 191 & 359 & 1000 
& $925^{+836}_{-488}$ & $1175^{+727}_{-457}$ & $353^{+12}_{-12}$\\
13 & 1.34 & 1.25 & 112 & 424 & 1150 
& $463^{+287}_{-159}$ & $1099^{+249}_{-233}$ & $424^{+12}_{-12}$\\
14 & 1.41 & 1.40 & 208 & 255 & 739 
& $894^{+846}_{-504}$ & $1024^{+770}_{-455}$ & $243^{+12}_{-16}$\\
15 & 1.36 & 1.25 & 204 & 405 & 1110 
& $1106^{+894}_{-596}$ & $1309^{+824}_{-532}$ & $395^{+17}_{-16}$\\
16 & 1.49 & 1.46 & 130 & 187 & 558 
& $286^{+259}_{-132}$ & $554^{+216}_{-149}$ & $177^{+7}_{-8}$\\
17 & 1.57 & 1.31 & 186 & 225 & 648 
& $705^{+678}_{-396}$ & $850^{+625}_{-351}$ & $224^{+9}_{-9}$\\
18 & 1.54 & 1.45 & 198 & 171 & 515 
& $645^{+656}_{-374}$ & $740^{+604}_{-325}$ & $167^{+10}_{-10}$\\
19 & 1.33 & 1.22 & 153 & 466 & 1250 
& $733^{+616}_{-341}$ & $1257^{+531}_{-385}$ & $472^{+11}_{-12}$\\
20 & 1.44 & 1.29 & 205 & 309 & 870 
& $962^{+843}_{-539}$ & $1134^{+803}_{-479}$ & $312^{+10}_{-13}$\\
\hline
\hline
\end{tabular*}
\caption{Source parameters for the $20$ BNS 
considered in this work. The last three columns provide the 
$68\%$ confidence intervals around the median for the 
tidal deformability measured by the LIGO-Virgo network~(HLV) and ET for each event.}
\label{table_source}
\end{table*}
%\end{comment}

\bibliographystyle{utphys}
\bibliography{Ref}

\end{document}